\documentclass{cccg05}
\usepackage{graphicx,amssymb,amsmath}
\usepackage{subfigure}
\usepackage{epsfig}

\newcommand{\stress}{\mbox{\it stress}}
\newcommand{\st}{\mbox{\it st}}
\newcommand{\Ar}{\mbox{\it Ar}}
\newcommand{\betw}{\mbox{\it betw}}

\title{A New Approach for Boundary Recognition in Geometric Sensor Networks}

\author{S\'andor P.~Fekete\thanks{Department of Mathematical Optimization,
        Braunschweig University of Technology, {\tt [s.fekete,a.kroeller@tu-bs.de}}
        \and
        Michael Kaufmann\thanks{Department of Computer Science, University of T\"ubingen, {\tt [mk,lehmannk@informatik.uni-tuebingen.de}}
        \and
        Alexander Kr\"oller\footnotemark[1]\ \thanks{Supported by the German Research Foundation (DFG) within the focus program ``Algorithms for Large and Complex Networks'' (SPP 1126), grant Fe407/8-1.}
        \and
        Katharina Lehmann\footnotemark[2]\ \thanks{Supported by the German Research Foundation (DFG) within the focus program ``Algorithms for Large and Complex
Networks'' (SPP 1126), grant Ka812/11-1.}
}
        
\index{Fekete, S\'andor P.}
\index{Kaufmann, Michael}
\index{Kr\"oller, Alexander}
\index{Lehmann, Katharina}

\begin{document}
\maketitle

\begin{abstract}
We describe a new approach for dealing with the following central
problem in the self-organization of a geometric sensor network: 
Given a polygonal region $R$, and a large, dense set of sensor nodes that are scattered 
uniformly at random in $R$. There is no central control unit, and nodes can only communicate locally by 
wireless radio to all other nodes that are within communication radius $r$, 
without knowing their coordinates or distances to other nodes.
The objective is to develop a simple distributed protocol that allows
nodes to identify themselves as being located near the boundary of $R$
and form connected pieces of the boundary.
We give a comparison of several centrality measures commonly
used in the analysis of social networks and show that 
{\em restricted stress centrality} is particularly
suited for geometric networks; we provide mathematical as
well as experimental evidence for the quality of this measure.
\end{abstract}

\section{Introduction}
\label{sec:intro}

In recent time, the study of wireless sensor networks (WSN) has become
a rapidly developing research area that offers fascinating
perspectives for combining technical progress with new applications of
distributed computing. Typical scenarios involve a large swarm of
small and inexpensive processor nodes, each with limited computing and
communication resources, that are distributed in some geometric
region; communication is performed by wireless radio with limited
range.  As energy consumption is a limiting factor for the lifetime of
a node, communication has to be minimized. Upon start-up, the swarm
forms a decentralized and self-organizing network that surveys the
region.

\begin{figure}[t]
\begin{center}
  \centering
  \subfigure[60,000 sensor nodes, distributed uniformly at random in a polygonal region.\label{fig:city:b}]{
    \epsfig{file=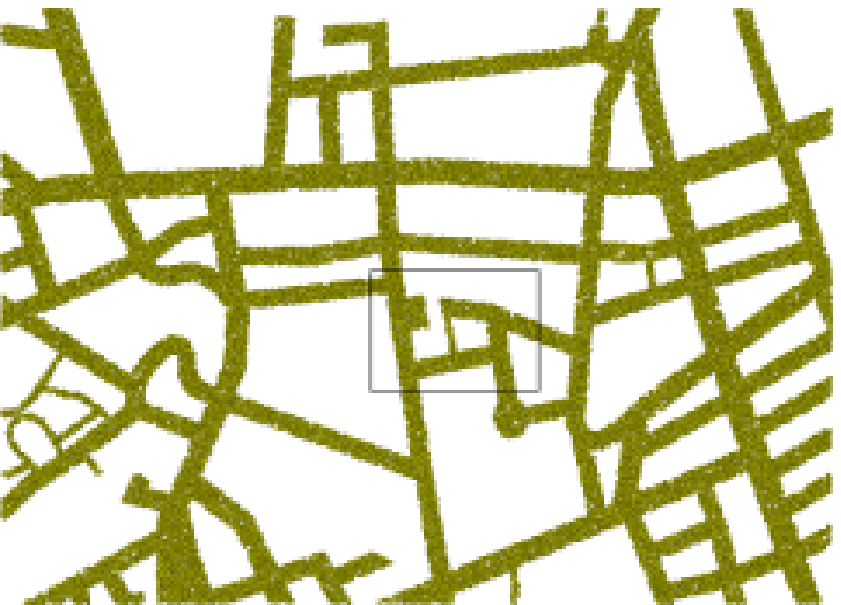,width=0.70\columnwidth}
  }
  \subfigure[A zoom into (a) shows the communication graph.\label{fig:city:c}]{
    \epsfig{file=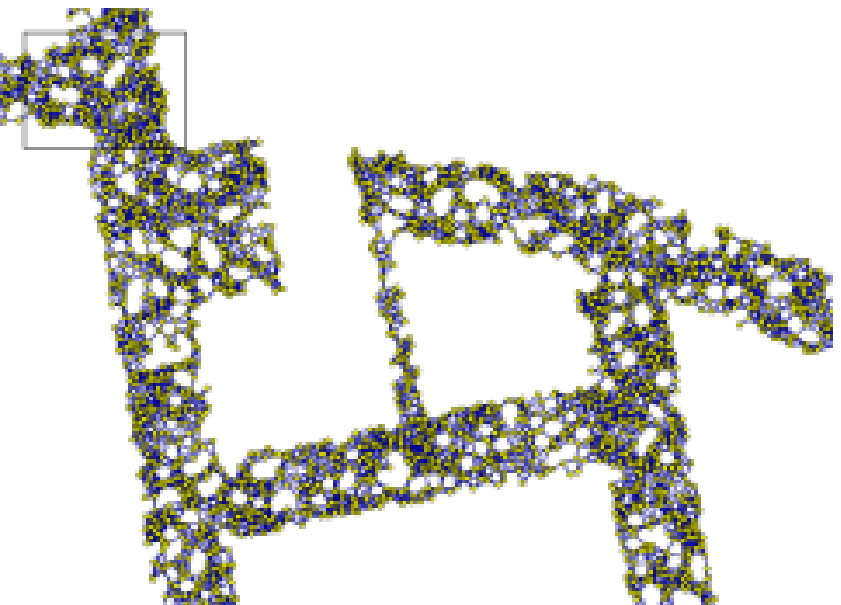,width=0.35\columnwidth}
  }
  \subfigure[A further zoom into (b) shows the communication ranges.\label{fig:city:d}]{
    \epsfig{file=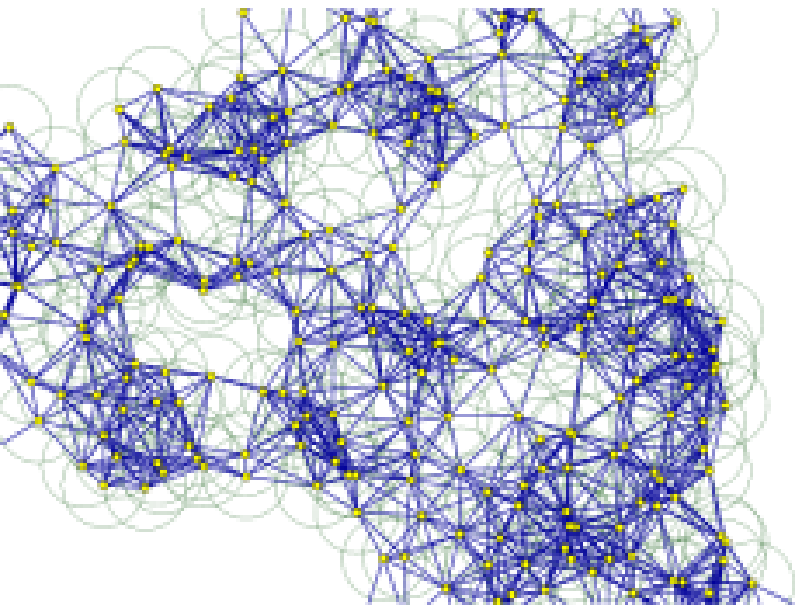,width=0.35\columnwidth}
  }
  \vspace*{-6mm}
  \caption{Scenario of a geometric sensor network, obtained by scattering sensor nodes in the street network surrounding Braunschweig University of Technology.}
  \label{fig:city}
\end{center}
  \vspace*{-6mm}
\end{figure}

From an algorithmic point of view, the characteristics of a sensor
network require working under a paradigm that is different from
classical models of computation: Absence of a central control unit,
limited capabilities of nodes, and limited communication between nodes
require developing new algorithmic ideas that combine methods of
distributed computing and network protocols with traditional
centralized network algorithms. In other words: How can we use a
limited amount of strictly local information in order to achieve
distributed knowledge of global network properties?

This task is much simpler if the exact
location of each node is known. Computing node coordinates 
has received a considerable amount of attention.
Unfortunately, computing exact coordinates requires the use of
special location hardware like GPS, or alternatively, 
scanning devices, imposing physical demands on size and structure
of sensor nodes.   As we demonstrated in our paper~\cite{kfb-kl-05},
current methods for computing coordinates based on anchor points
and distance estimates encounter serious
difficulties in the presence of even small inaccuracies, which are
unavoidable in practice.

As shown in \cite{fkp-nbtrsn-04}, there is a way to sidestep many of the above
difficulties, as some structural location aspects do {\em not}
depend on coordinates. 
This is particularly relevant for sensor networks 
that are deployed in an environment
with interesting geometric features. (See \cite{fkp-nbtrsn-04}
for a more detailed discussion.) Obviously, scenarios as the one
shown in Figure~1 pose a number of interesting
geometric questions. Conversely, exploiting the basic fact
that the communication graph of a sensor network 
has a number of geometric properties provides
an elegant way to extract structural information.

One key aspect of location awareness is {\em boundary recognition},
making sensors close to the boundary of the surveyed region 
aware of their position and letting them form
connected {\em boundary strips} along each verge.  
This is of major importance for keeping track of events entering or
leaving the region, as well as for communication with the
outside.  Neglecting the existence of holes in the region may also
cause problems in communication, as routing along shortest paths tends
to put an increased load on nodes along boundaries, exhausting their
energy supply prematurely; thus, a moderately-sized hole (caused by
obstacles, by an event, or by a cluster of failed nodes) may tend to
grow larger and larger.

We show that using a combination of geometry, stochastics, and tools
from social networks, a considerable amount of location awareness can indeed be
achieved in a large swarm of sensor nodes without any use of location
hardware. The result is a relatively simple distributed algorithm 
for boundary recognition in large geometric sensor networks that shows
excellent performance for test networks with 80,000 nodes.

\section{Centrality Measures for Social Networks}
\label{social}
A different area studying large and complex graphs is the field
of {\em Social Networks}, where nodes represent individuals
in a large collective, and edges indicate some interaction between
them. (See the recent book \cite{be-namf-05} for an overview and an extensive
list of references.) Identifying asymmetries within a network
is a natural approach; one particular way of doing this is based
on so-called centrality indices, i.e., real-valued functions that 
assign high values to more ``central'' nodes, while ``boundary'' nodes
get low values. 

In the last five decades, many different centrality 
indices have been proposed. There are two major classes: One is based
on local properties of the graph, so it is particularly suited for
typical scenarios of sensor networks and will be discussed in some detail. 
The other class is based on more global properties, e.g., 
the computation of eigenvalues of the adjacency matrix, so it is less
useful for our purposes. 

Centrality indices of the first class can be subdivided into three 
subclasses: The first considers the distances to other vertices,
the second determines the number of vertices at a given 
distance, while the third makes use of shortest 
paths containing a given vertex. 

\begin{figure}
\begin{center}
  \centering
  \includegraphics[width=0.6\columnwidth]{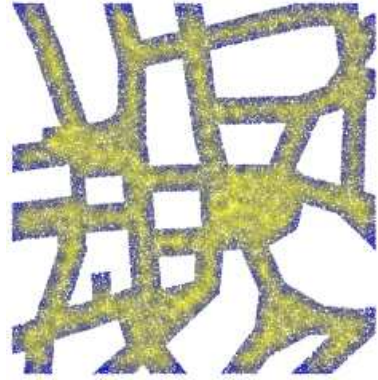}
  \caption{$k$-hop neighborhood for $k$=4.}
  \label{fig:cent:a}
\end{center}
  \vspace*{-6mm}
\end{figure}

Considering the maximum distance to another vertex in the graph
(based on hop-count) does not reflect local topological structures
in a sensor network; in particular, it fails to indicate closeness
to interior boundaries. The size of the $k$-hop neighborhood
is better suited, and (for the simple choice $k=1$) was indeed the basis
for our approach described in \cite{fkp-nbtrsn-04}, as it is an indicator
for the size of the intersection of the communication range of
a node with $R$.
It is tempting to try to improve the results by increasing $k$,
but this is not without drawbacks with respect to topological properties, 
as a boundary node
close to a ``thick'' part of $R$ may get a better value
than an interior node that is located in a ``thin'' part of the region.
See Figure~\ref{fig:cent:a} for a scenario with 80,000 nodes;
index values are represented on a color scale from dark (low)
to light (high).

This leaves the structure of shortest paths. In particular,
the {\it stress centrality} $stress(v)$ is defined as the number 
of shortest paths containing $v$:
\begin{equation}
\stress(v) := \sum_{s \in V}\sum_{t \not = s \in V} \sigma_{st}(v),
\end{equation}
where $\sigma_{st}(v)$ denotes the number of shortest paths containing $v$. 
Only considering vertices within a given distance $\delta$ yields 
the {\em restricted stress centrality}:
\begin{equation}
\stress(v, \delta) := \sum_{s \in V_\delta(v)}\sum_{t \not = s \in V_\delta(v)} \sigma_{st}(v).
\end{equation}

\begin{figure*}
\begin{center}
  \subfigure[Betweenness centrality.\label{fig:cent:b}]{
    \epsfig{file=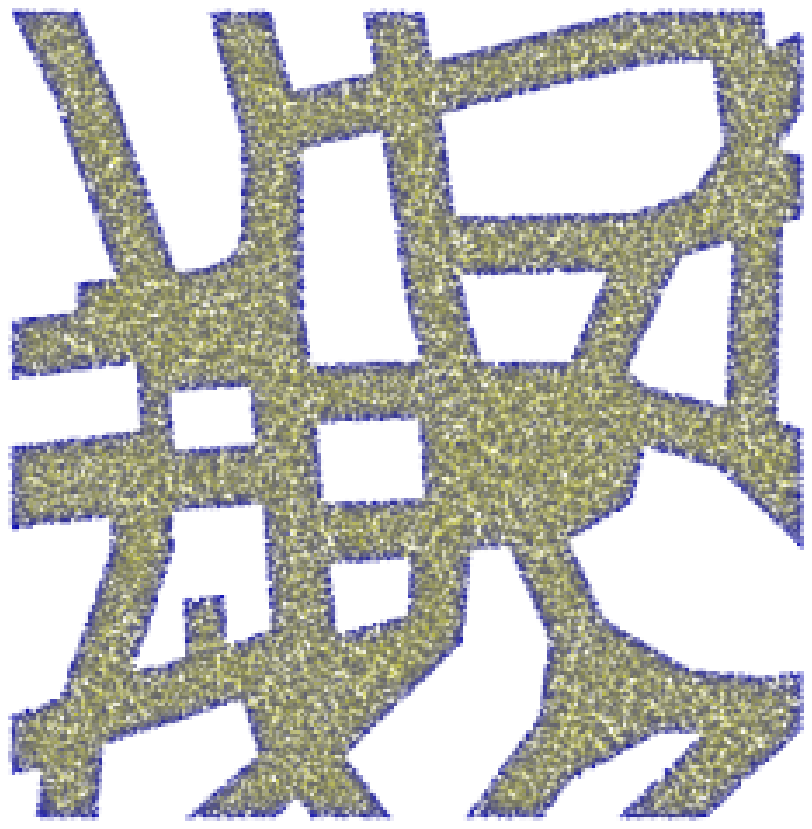,width=0.65\columnwidth}
  }
  \subfigure[Stress centrality.\label{fig:cent:c}]{
    \epsfig{file=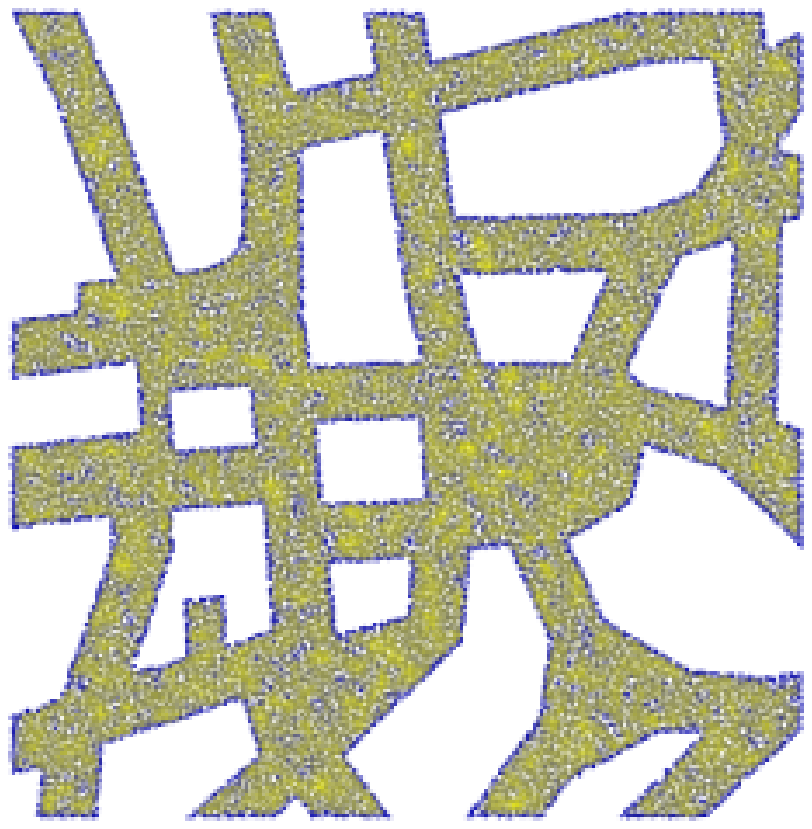,width=0.65\columnwidth}
  }
  \subfigure[Restricted stress centrality with threshold filter.\label{fig:cent:d}]{
    \epsfig{file=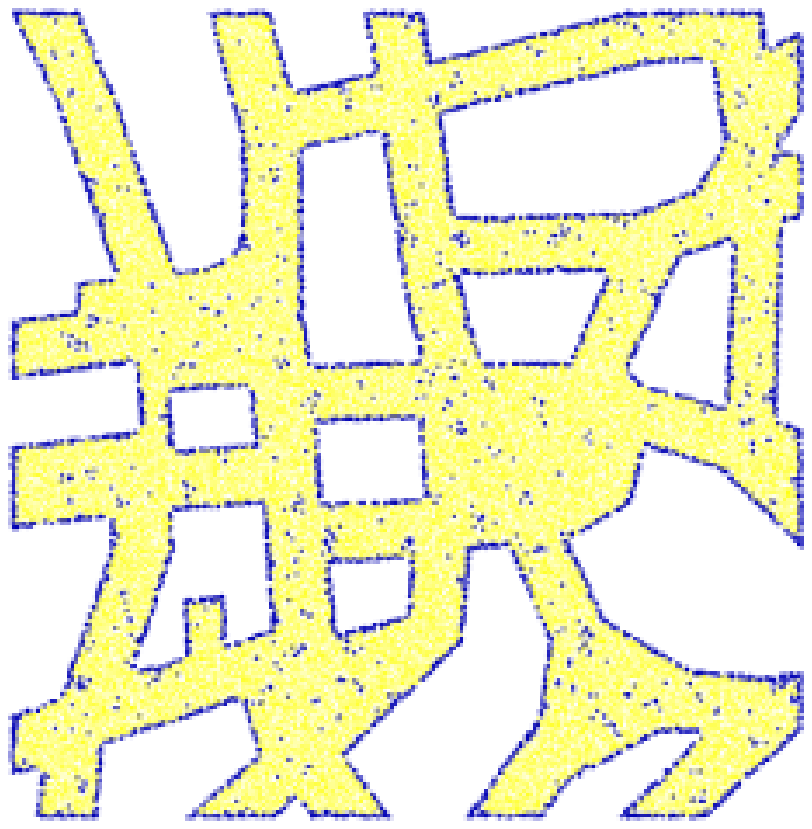,width=0.65\columnwidth}
  }
  \vspace*{-6mm}
  \caption{Performance of different centrality measures, shown for a scenario of 80,000 nodes distributed uniformly at random.}
  \label{fig:perform}
\end{center}
  \vspace*{-6mm}
\end{figure*}

In the context of a communication network, this measure can be 
motivated as follows: 
If each vertex sends a message to every other vertex along all shortest paths, 
the stress centrality counts how many times vertex $v$ is busy with 
passing on a message. As there may be quite many shortest paths,
it is reasonable to assume that a vertex
sends a message to some other vertex and uses any of their shortest paths with the same probability, i.e., $1/\sigma_{st}$, where 
$\sigma_{st}$ denotes the number of shortest paths between $s$ and $t$.
The probability of any vertex $v$ that it has to transport the message is thus 
given by $\rho_{st}(v):=\frac{\sigma_{st}(v)}{\sigma_{st}}$.
The {\it betweenness centrality} $\betw(v)$ is defined as the sum over all $\rho_{st}(v)$:
\begin{equation}
\betw(v) := \sum_{s \in V}\sum_{t \in V} \rho_{st}(v).
\end{equation}
See Figure~\ref{fig:cent:b} for the evaluation of betweenness centrality
for our example, while Figure~\ref{fig:cent:c} shows the stress centrality.
(Again, low values are indicated by dark dots, while high values are represented
by light color.)
A detailed analysis for restricted stress centrality 
is given in the following section.

%
%

\section{Using Restricted Stress Centrality}
\label{stress}
In the context of a sensor network, it takes a number of algorithmic
steps to evaluate a measure and use the results for extracting
global features like boundaries. Some of those details are described
in our paper \cite{fkp-nbtrsn-04}, and can be used analogously for
other measures: Using an auxiliary tree structure (which is easy
to obtain), we can aggregate local results globally in order
to determine appropriate threshold values. Once a threshold has been set,
it can be distributed to all nodes in the network; after that, each
node simply checks whether its centrality index is above or below
the threshold, resulting in a classification as ``interior'' or ``boundary''.
A good index must have the following properties:
\begin{itemize}
\item It should require only simple local computations for each node.
\item Setting a good threshold value should be relatively easy.
In other words: The distributions for interior nodes and for boundary nodes
should be well-separated.
\end{itemize}

\begin{theorem}
\label{th:sep}
Using the restricted stress centrality $\stress(v,1)$, 
nodes are classified correctly with high probability
for sufficiently large node density.
\end{theorem}

See Figure~\ref{fig:cent:d} for the result for restricted stress centrality
for relatively moderate density:
It can be seen that all boundary nodes are correctly classified. The
interior contains a number of false positives, which can be eliminated 
by additional filters.

\begin{figure}[h]
  \centering
  \includegraphics[width=.45\columnwidth]{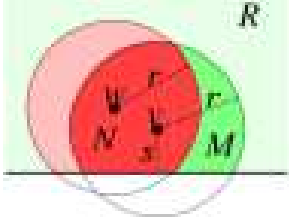}
  \vspace*{-3mm}
  \caption{For any given neighbor $w$ of $v$, the expected fraction of 
  neighbors of $v$ that are not neighbors of $w$ is given by 
  $\frac{|M|}{|N\cup M|}$.}
  \label{fig:circles}
  \vspace*{-3mm}
\end{figure}

{\bf Discussion of Theorem~1.}
Let $v$ be a node in the network, and let $\delta(v)$ be the number
of neighbors of $v$. Furthermore, $\stress(v,1)$ is the number
of nonadjacent neighbors of $v$. Then the normalized
coefficient $\st(v):=\frac{2\stress(v,1))}{\delta(v)(\delta(v)-1)}$
describes the fraction of pairs of neighbors that are nonadjacent,
i.e., that have a shortest-path connection via $v$, so 
$\mathbb{E}[\stress(v,1)]=\mathbb{E}[st(v)]\left(\begin{array}{c}{\mathbb{E}[\delta(v)]}\\2\end{array}\right)$. 
Now consider any neighbor $w$ of $v$. Let $C(v):=\{p\in R\mid d(p,v)\leq r\}$
be the portion of $R$ that is within communication range of $v$.
See Figure~\ref{fig:circles}; let $N_w:=C(v)\cap C(w)$, and
$M_w:=C(v)\setminus C(w)$. For a uniform random distribution,
the expected fraction of neighbors of $v$ that are not adjacent
to $w$ corresponds to the ratio of areas 
$\frac{\Ar(M_w)}{\Ar(C(v))}$.
Integrating over all possible positions of $w$, we get
an overall expected value 
$\st(v)=\frac{1}{\Ar(C(v))}\int_{w\in C(v)}\left(\frac{\Ar(M_w)}{\Ar(C(v))}\right)dw$.

As the size of the areas also depends on the distance $s$ of $v$
from the boundary, solving this integral in closed form 
for all $s$ would require finding a primitive that contains $d$ as an explicit
parameter; this appears to be hopeless, even using ideas as described
in \cite{geo.prob}. However, for specific values of $s$,
an explicit numerical calculation is possible:
For $s\geq r=1$ and $d(w,v)=x$ the area of $M_w$ turns out to be
$\frac{8\left(\arccos\left(\frac{x}{2}\right)-\frac{1}{2}\sin\left(2\arccos\left(\frac{x}{2}\right)\right)\right)}{3}$. 
The resulting integral $\sigma=\int_0^1 x\left(1-\frac{2\left(\arccos\left(\frac{x}{2}\right)-\frac{1}{2}\sin\left(2\arccos\left(\frac{x}{2}\right)\right)\right)}{\pi}\right)dx$ can be solved numerically,
resulting in a value of $\sigma=0.4134966716$. 

\begin{figure}
\begin{center}
  \centering
  \subfigure[Distributions for neighborhood size 20.\label{fig:dist:a}]{
    \epsfig{file=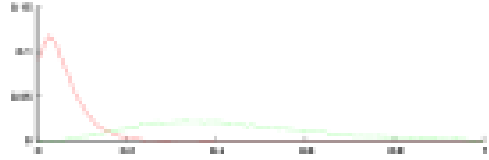,width=0.8\columnwidth}
  }
  \subfigure[Distributions for neighborhood size 200.\label{fig:dist:b}]{
    \epsfig{file=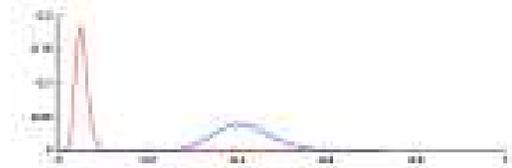,width=0.80\columnwidth}
  }
  \vspace*{-3mm}
  \caption{Random distribution of restricted stress centrality for a node on the boundary and in the interior,
for different neighborhood sizes.}
  \label{fig:dist}
\end{center}
  \vspace*{-3mm}
\end{figure}

For determining threshold values for separating interior and
boundary values of $\st$, we also need the random distribution
of $\st$ for different values of $s$. These distributions
can be determined with additional numerical computations; using
a Monte-Carlo simulation, we obtained distributions
like the ones in Figure~\ref{fig:dist}: Shown are the distributions
for 20 expected neighbors (\ref{fig:dist:a})
and for 200 expected neighbors (\ref{fig:dist:b}); the left
(red) curve shows the distribution of $\st$ for a node $v$ on the
boundary, while the right (green/blue) curve shows the distribution
completely in the interior of $R$.
The probability of error for a specific threshold is given 
by the normalized area to the right of the threshold below
the left curve (false negatives)
or by the normalized area to the left of the threshold below
the right curve (false positive). Clearly, the error becomes
arbitrarily small for large neighborhood size. 
\QED

For intermediate sizes
as the one in our example, choosing a relatively large threshold
value avoids too many false negatives, at the expense of a limited
ratio of false positives.

\section{Algorithm}

In \cite{fkp-nbtrsn-04}, we showed how to estimate
$\mathbb{E}[\delta(v)]$ for a node $v$ of boundary distance $s\geq r$,
i.e., a node on the inside of the network. The algorithm constructs a
tree, collects a node degree histogram and floods the result to all
nodes. Both the total runtime of the algorithm and the total size of
messages is $\mathcal{O}(|V|\log^2|V|)$. Each node stores a constant
threshold value
$0 < \theta < \sigma$ that has been chosen in advance. If
\[ st(v)\leq \theta\left(\begin{array}{c}\mathbb{E}[\delta(v)]\\2\end{array}\right) \;, \]
the node declares itself to be a boundary node. In experiments, we
found $\theta=1/3$ to be a particularly good choice.

\section{Conclusion}

We showed that restricted stress centrality is a useful index
for extracting topological boundary information from a geometric
sensor network, provided that the distribution of nodes follows
a suitable random distribution. As this is a rather strong assumption,
it appears desirable to come up with more general methods.
Moreover, an approach based on random distributions
may still fail in some rare cases
(even though the probability of failure is extremely low),
so it is particularly interesting to develop
deterministic methods for boundary recognition.
Such an approach is described in our forthcoming paper
\cite{fkfp-dbrlgsn-05}.



\small 
\bibliographystyle{abbrv}
\bibliography{refs}


\end{document}